\documentclass[12pt,preprint]{aastex} 
\newcommand\etal{{\it et al.}}
\newcommand\lae{\mathrel{<\kern-1.0em\lower0.9ex\hbox{$\sim$}}}
\newcommand\gae{\mathrel{>\kern-1.0em\lower0.9ex\hbox{$\sim$}}}
\newcommand\kms{km~s$^{-1}$}
\newcommand\mone{$^{-1}$}
\newcommand\mtwo{$^{-2}$}
\newcommand\mthree{$^{-3}$}

\begin{document}

\title{ XMM-NEWTON Detection of X-ray Emission from the Compact Steep Spectrum
Radio Galaxy 3C~303.1   }
\author{Christopher P. O'Dea\altaffilmark{1}, Bo Mu\altaffilmark{2},
D. M. Worrall\altaffilmark{3}, Joel Kastner\altaffilmark{2}, 
Stefi Baum\altaffilmark{2}, and   Willem H. De Vries\altaffilmark{4}} 

\altaffiltext{1}{Dept of Physics, RIT }
\altaffiltext{2}{Center for Imaging Science, RIT }
\altaffiltext{3}{University of Bristol}
\altaffiltext{4}{IGPP/LLNL }


\begin{abstract}
Using XMM we detect faint unresolved X-ray emission from the Compact 
Steep Spectrum radio galaxy
3C~303.1. We detect a thermal component at $kT \simeq 0.8$ keV which seems likely
to be produced in the ISM of the host galaxy. There is evidence for a second
component in the spectrum whose nature is currently ambiguous. Plausible
hypotheses for the second component include (1) hot gas shocked by the expansion
of the radio source, and (2) Synchrotron self-Compton emission from the southern 
radio lobe if the magnetic field is below the equipartition value by a factor 
of $\sim 3.5$.  
\end{abstract}


\section{INTRODUCTION}

The progenitors of the large-scale powerful classical double (3CR FRII)
sources seem likely to be the GHz Peaked Spectrum (GPS) and Compact
Steep Spectrum (CSS) radio sources
(e.g., O'Dea, Baum, \& Stanghellini 1991; Fanti \etal\ 1990;
Fanti \etal\ 1995; Readhead \etal\ 1996a,b; O'Dea \&
Baum 1997, for a review see O'Dea 1998).   The GPS and CSS sources 
have generally simple and convex radio spectra with peaks near 
1 GHz and 100 MHz, respectively. 

Models for the evolution of powerful radio galaxies
(Fanti \etal\ 1995; Begelman 1996;
Readhead \etal\ 1996b; O'Dea \& Baum 1997; De Young 1997;
Kaiser \& Alexander 1997; Kaiser, Dennett-Thorpe, \& Alexander 1997;
O'Dea \& Baum 1997; Blundell, Rawlings \& Willott 1999; 
and cf Snellen \etal\ 2000) 
are generally consistent with a scenario
in which  these sources propagate from the $\sim 10$ pc to Mpc scales
at roughly constant velocity through
an ambient medium which declines in density 
while the sources decline in radio luminosity.
 

There is increasing evidence that the GPS and CSS sources
interact with the host galaxy as they propagate through it. 
Broad- and narrow-band HST imaging of CSS radio sources
(De Vries \etal\ 1997, 1999; Axon \etal\ 2000) have shown
that CSS radio galaxies at all redshifts exhibit emission-line gas
which is strongly aligned with the radio source.
The alignment is much stronger than seen in low redshift
large-scale radio galaxies and is similar to that seen in high
redshift ($z \gae 0.6$) radio galaxies. 
The close association between the gas and the radio source suggests
that the latter interacts strongly with the ambient medium as
it propagates through the ISM.

In 3C~303.1 cooling time arguments (e.g., Pedlar, Dyson, \& Unger 1985; 
Taylor, Dyson, \& Axon 1992)  applied to the gap between the radio hot 
spots and the onset of bright emission-line gas are
consistent with lobe expansion velocities $\gae 6000$ \kms\ for ambient
densities of $n \sim 1$ cm\mthree\ (De Vries \etal\ 1999).  The broad
and highly structured spatially integrated [OIII]$\lambda 5007$
line widths observed by Gelderman \& Whittle (1994) strongly suggest
that the radio source is dominating the emission-line kinematics.
HST/STIS long-slit medium-dispersion spectroscopy of the
aligned emission-line gas has shown that the kinematics of the gas
are consistent with cloud motions driven by  interaction with the expanding
radio source (O'Dea \etal\ 2002).  Constraints on the bow shock velocity
from lifetime and cooling arguments are in the range few 
$\times 10^3$ km s\mone\ to few $\times 10^4$ km s\mone\
(O'Dea \etal\ 2002). Analysis of the low dispersion STIS
spectra reveals that the ionization diagnostics are consistent with
strong contributions from shocked gas (Labiano \etal\ 2005).
HST UV imaging shows a clear detection of UV continuum light tightly 
aligned with the radio
source (Labiano \etal\ 2006). These observations are all consistent with
the radio source in 3C~303.1 interacting with gas clouds in its environment,
shocking them and triggering star formation. 

The interaction of the radio source with its environment may also produce 
observable signatures in the X-ray.  Heinz, Reynolds \& Begelman (1998)
suggested that compact evolving radio galaxies might create shells of
hot shocked gas as they expand into the ambient medium. 
X-ray observations have found examples of both hotter (e.g., 
Kraft \etal\ 2003; Wilson, Smith \& Young 2006) and cooler 
(e.g., Nulsen \etal\ 2005; Soker, Blanton \& 
Sarazin 2002; Blanton \etal\ 2001) shells of ambient gas swept up by expanding
radio galaxies. 
  
In this paper we present XMM-Newton X-ray observations of the CSS radio 
galaxy 3C~303.1. We discuss the results and their implications for
our understanding of CSS radio sources. 

\section{THE XMM OBSERVATIONS AND RESULTS}
\label{sec:obs}

The properties of the source are given in Table 1. 
We observed 3C~303.1 with XMM-Newton for about 40~ks on August, 18, 2003. 
Three EPIC cameras, MOS-1, MOS-2 and pn, were operated with medium optical 
blocking filters in ``Prime Full Window" mode, which covers the full field-of-view 
(FOV) of 30' diameter. 
Because a fraction of the flux collected by the telescopes hosting the
MOS cameras is directed to dispersing grating arrays, the photons collected
by each MOS camera are fewer than those collected by the pn camera.

\begin{figure}
\plotone{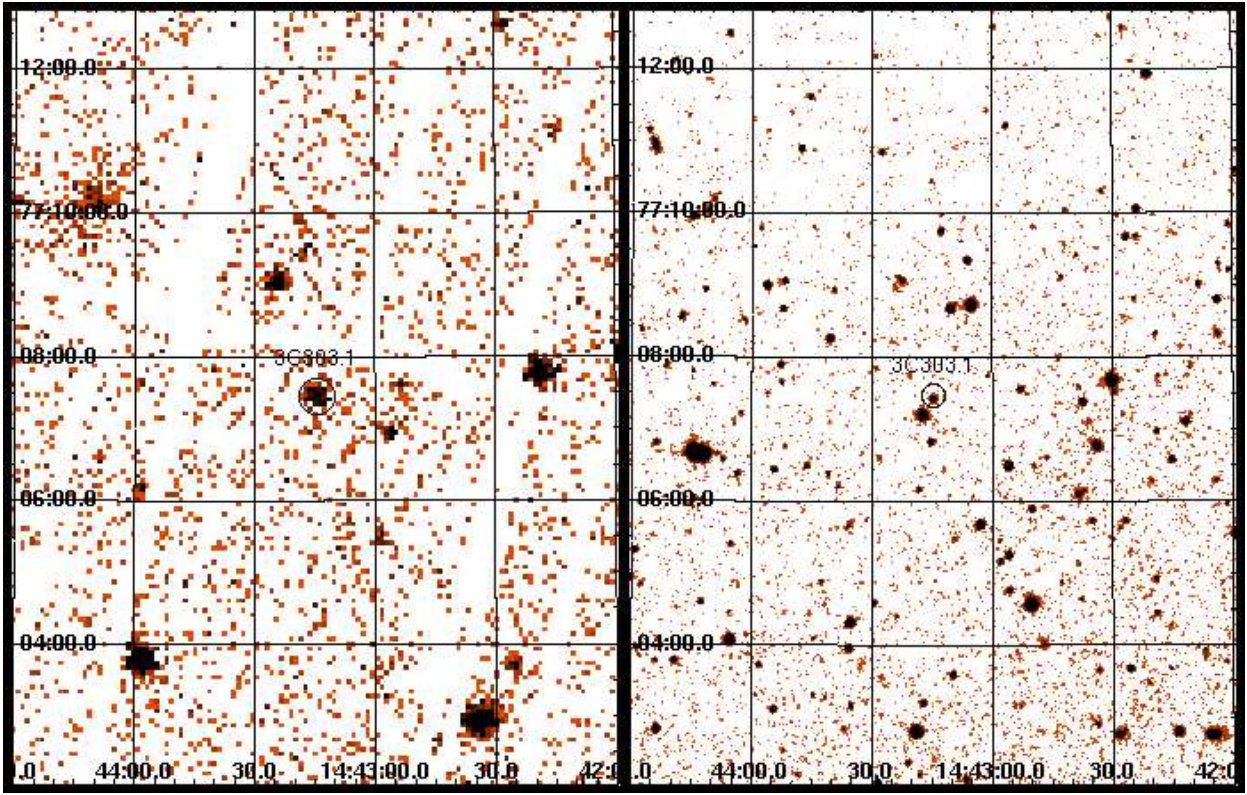} 
\caption{3C~303.1. (Left). XMM pn image showing detection of 3C303.1.
A circle of 15 arcsec radius is shown around the source. 
(Right). Optical image from the STScI digitized sky survey. A circle of 10 
arcsec radius is shown around the source.
}
\label{image}
\end{figure}

The data were reprocessed using the XMM-Newton Science Analysis System (SAS) 
v6.1.0. EMCHAIN and EPCHAIN were employed to obtain the photon event lists. 
Then we used EVSELECT to select single, double, triple and quadruple 
(PATTERN $\lae 12$) for MOS, and single and double events (PATTERN $\lae 4$) 
for pn with the energy range from 0.2 keV to 12 keV.  FLAG was set to zero 
to reject the events which are close to the CCD gap and bad pixels. 
Because of the high EPIC background due to contamination 
by solar protons we restricted the data using Good Time Interval (GTI) filters.
Details of the resultant exposure times and count rates in the three cameras
are given in Table 2. The pn image, showing the detection of a faint source at the
location of 3C~303.1 is given in the left-hand panel of Figure~\ref{image}.

\subsection{Source Size}

The radial profiles of the source counts are consistent with the X-ray source being 
unresolved by the pn and MOS detectors. This is consistent with the emission being
produced by the radio source (1.7 arcsec) and/or the host
galaxy (20 kpc = 4.9 arcsec). Galaxy counts by Harvanek \etal\ (2001) show no
evidence for a cluster of galaxies surrounding 3C~303.1.  

\subsection{Spectral Fitting}

We chose an on-source extraction circle of 40 arc-second radius, which at 1.5 keV 
corresponds to $\sim 87\%$ encircled energy for the MOS and 
$\sim 83\%$ encircled energy for the pn. 
The MOS local background region was an annulus
 surrounding 3C~303.1 with inner and outer radii of 40 and 80
arcsec, respectively, with two unrelated sources masked.
For the pn the background region was
a circle the same size as the on-source region but
from a source-free region of the same RAWY.
For each camera the task ESPECGET was employed to
generate a background-subtracted spectrum, grouped with 25 counts minimum per 
spectral bin to improve the statistics, a Redistribution Matrix File 
(RMF), and an Auxiliary Response File (ARF).
Data above 8 keV and below 0.4 keV were excluded
from the spectral model fitting due to the poor statistics at these energies. 
The resulting count rates are given in Table 2.

\begin{figure}
\epsscale{0.80}
\plotone{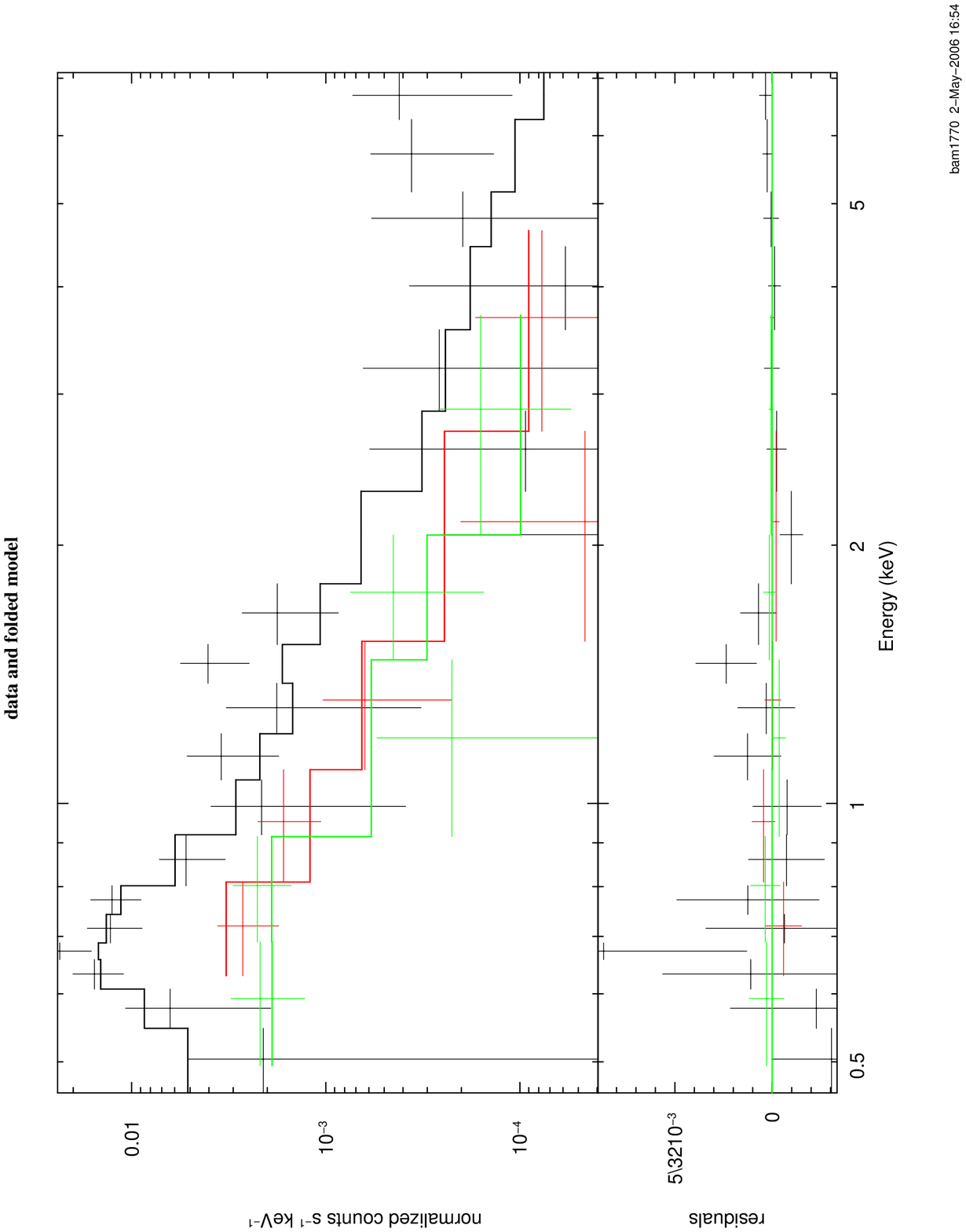}
\caption{The XMM spectrum of 3C303.1 showing the pn (darker) and MOS (lighter) data and the model
of Table~\ref{tabthermpow}. 
}
\label{spectra1}
\end{figure}

We fitted the pn and MOS spectra simultaneously with several 
models consisting of different combinations of absorption with power law (PL)
and Raymond-Smith (RS) thermal models using XSPEC 12.2.0cf.  
The results are given in Tables 3--5.
We obtain an acceptable fit for a RS model with Galactic absorption (frozen),
solar abundance (frozen), and a temperature of kT$\sim 0.8$ keV (Table~\ref{tabtherm1}).
We do not obtain an acceptable fit with a single PL model with either
Galactic absorption alone or excess intrinsic absorption. 
Our single RS fit improves if we add a PL component (Table 4), 
though the photon index is very poorly constrained by our data. 
Assuming a plausible photon index, e.g., $\Gamma = 1.5$, there is no need 
for absorption in excess of Galactic towards the PL component. 
In order to constrain a possible hot component due to gas shocked by
the radio source (\S \ref{sechotgas}) we considered a two temperature
model with $T_1 = 0.8$ and $T_2 = 45$ keV (Table 5). Such a model
is also an improvement
over a single temperature thermal model.  
Figure ~\ref{spectra1} shows the spectral data together with the model
components of
Table~\ref{tabthermpow} folded through the detector response files.

\section{DISCUSSION}\label{sec:impl}

\subsection{The Thermal Component}

The luminosity ($L_x \sim 2\times 10^{42}$ ergs s\mone) and temperature 
(kT$\sim 0.8$ keV) of the RS component are consistent with the 
values found in elliptical galaxies hosting radio sources 
(e.g., Worrall \& Birkinshaw 
2000) and possibly small groups (e.g., Helsdon \& Ponman 2000). 
However, the X-ray emission is both too faint and
too cool to be consistent with a cluster of galaxies (e.g.,
Arnaud \& Evrard 1999). This suggests that we have detected the ISM
of the host galaxy or possibly the intragroup medium (if there is a
galaxy group). 

\subsection{Constraints on shock heated gas\label{sechotgas}}

The expanding radio source will run over and shock the ambient gas possibly
producing detectable emission (e.g., Heinz \etal\ 1998). We calculate the 
expected properties of the shocked gas following 
Worrall \etal\ (2005). The sound speed in the 0.8 keV ISM is
$c_s \simeq 460$ km s\mone. If the propagation velocity of the radio source
is as high as 6000 km s\mone\ as suggested by the cooling-time arguments
for the emission line nebula (De Vries \etal\ 1999), 
the Mach number is M $\simeq 13$. For such a strong
shock the Rankine-Hugoniot conditions give a temperature ratio
between shocked and pre-shocked gas of $T_2/T_1 \simeq
54$, which implies $T_2 \simeq 43$ keV. We note that a two-temperature model with 
$T_1 = 0.8$ and $T_2 = 45$ keV is permitted by the data (Table~\ref{tabt2}).  

Taking into account the uncertainties, Table~\ref{tabt2} shows that 
the normalizations
of the heated (shocked) to cool (unshocked) gas is a factor of roughly 0.9 to 6.
The 1-keV normalization is proportional to $n^2 V$, where $n$ is the number density
of electrons or protons and $V$ is the volume.
Since the Rankine-Hugoniot conditions for a strong shock give
$n_{\rm shocked}/n_{\rm unshocked} = 4$, the X-ray spectral results are
consistent with $V_{\rm unshocked}/V_{\rm shocked} \sim 3$ to 18.
What is the volume of the shocked gas?
We detect radio emission from the cocoon but not from the hot shocked
gas which lies between the cocoon and the bow shock. We use the numerical simulations 
of Carvalho \& O'Dea (2002) as a guide
to estimate the ratio of the volumes of hot shocked gas and cocoon. 
The estimated ratio is model dependent and varies with the jet density
contrast and Mach number. For a light, supersonic jet the numerical simulations
suggest a value of about 3 for the ratio of hot shocked gas to cocoon volume.
Adopting a value of 3 along with the estimates of cocoon volume from
the MERLIN observations gives a volume of shocked gas of about 14 kpc$^3$.  
We find
the volume of unshocked gas is large enough to be consistent
with a realistic galaxy atmosphere if $V_{\rm unshocked}/V_{\rm shocked}$
is towards the upper value permitted by the X-ray spectral fitting.
For example, if $V_{\rm unshocked}/V_{\rm shocked}$
is 18, a spherical galaxy of uniform density would be required to have
a radius of about 4 kpc.
Such a galaxy has a similar $n^2 V$ to one of the same central density
modeled with a $\beta$ model of $\beta = 2/3$ and core radius of $\sim 3$ kpc.
{\it Chandra\/} measurements of the atmospheres around
relatively powerful nearby radio galaxies are fitted to $\beta$ models that
typically bracket these parameter values (e.g., 
Kraft et al. 2005, Worrall \& Birkinshaw 2005).
High-resolution X-ray imaging spectroscopy would
probe the density and temperature structure and should confirm 
whether such a high-temperature component exists. 

\subsection{The Power-Law Component}

Here we consider the possibility that there is a significant non-thermal 
power-law component, i.e., the second component in the fit to the spectrum 
is not due to a hot shocked gas.
The lack of a high intrinsic absorbing column towards the PL component
suggests that it is not from 
accretion-disk emission that would be expected to be obscured in this
powerful two-sided radio galaxy, and the fact that the nucleus is undetected
both at 1.6 GHz (Sanghera \etal\ 1995)  and
at 5.0 GHz (L\"udke \etal\  1998) disfavors the X-ray emission
originating from a small-scale jet.
X-ray binaries produce emission that can mimic a hard power law, but
the luminosity of $> 10^{42}$ ergs s$^{-1}$ is one or two
orders of magnitude more than is typically measured in elliptical galaxies
(e.g., Kim \& Fabbiano 2004).

The most likely explanation for a power-law component would appear to
be inverse Compton emission from the lobes of the radio galaxy.
We estimate the expected flux
(e.g., Harris \& Grindlay 1979; Harris \& Krawczynski 2002;
Hardcastle, Birkinshaw \& Worrall 1998) 
using the MERLIN radio measurements of Sanghera \etal\ (1995), where at 1.6 GHz 
the Southern lobe contains about 90\% of the total radio 
flux density. 
We find that, because of the small size of the lobes, Synchrotron
self-Compton (SSC) dominates the X-ray yield from
scattering of the cosmic microwave background radiation.

Results of the SSC calculations are presented
in Table~\ref{tabssc}. 
In these calculations we adopt a single-component
power-law electron spectrum of 
slope consistent with the observed radio synchrotron spectrum
(Sanghera \etal). 
In order to  maximize the SSC output under the equipartition assumptions,
we set the minimum electron energy to no lower than is
required to produce the radio 
emission, and we assume no non-radiating particles (protons) are
present.  If the magnetic field is at 
the equipartition strength, the predicted SSC X-ray flux is too low by factors
of $\sim 350$ in the Northern lobe and $\sim 16$ in the Southern lobe.
However, if the field is below the equipartition level by factors of
$\sim 16$ and $\sim 3.5$, 
in the N and S lobes, respectively, the predicted flux is comparable
to that in the measured PL component. 
There is some evidence that the magnetic fields in 
some radio lobes are indeed about 1/3 of the equipartition strength 
(e.g., Carilli, Perley \& Harris 1994; Wellman, Daly \& Wan 1997; 
Croston \etal\ 2005), and thus there is additional justification for the PL
component arising from SSC emission primarily in the
S lobe of 3C~303.1, with a magnetic field about 1/3 of the equipartition
value. If this is the correct interpretation it suggests that
the mechanism for keeping the average B field below equipartition occurs
in young radio galaxies and is not just a feature of large FRIIs.

\subsection{Comparison with other Radio Galaxies }

There are now X-ray observations of a dozen GPS radio galaxies (O'Dea \etal\ 1996, 2000;
Iwasawa \etal\ 1999; Guainazzi \etal\ 2004, 2006; Vink \etal\ 2006).  4C+55.16
appears to be in a cluster with a cooling flow (Iwasawa \etal\ 1999). The others
are dominated by a power-law component with intrinsic absorption of
$N(H) \sim 10^{22}$ to
$ \sim 10^{24}$ cm\mtwo\ (Guainazzi \etal\ 2006; Vink \etal\
2006).  This intrinsic absorption is larger than the HI absorption
along the line of sight to the
radio emission in the small number of cases where the radio
measurements have been made (Vink \etal), thus
suggesting that the absorbed X-ray emission arises from the
vicinity of an accretion disk.  

Large narrow-emission-line FRII radio
galaxies generally have two nuclear X-ray components (Belsole et
al.~2006). One has
intrinsic absorption similar to that measured for most
GPS sources, and is interpreted as accretion related,
and the other is unabsorbed and identified as
arising from jet emission within the radio core.
Our X-ray measurements of the CSS radio galaxy 3C~303.1 find no evidence for
absorbed emission -- it is either absent or very weak.
If absent, it may suggest that the central engine of 3C~303.1 is
more similar to that of an FRI radio galaxy, where absorbed
X-ray components are rarely seen (Evans et al. 2006 -- but note
the cases of Cen~A [Evans et al.~2004] and NGC 4261 [Zezas et al.~2005]).
The radio core of 3C~303.1 is undetected, and if the second X-ray
spectral component is indeed non-thermal, then
it is much more plausibly associated with the radio lobes than with the
radio core.  The alternative hypothesis is that we are seeing shocked
gas.  In either case this hard emission should be spatially resolved
in {\it Chandra\/} observations.
It appears that the weak core of 3C~303.1 is an advantage in allowing
other X-ray components to be seen more clearly.

\section{SUMMARY}\label{sec:summary}

Using XMM we detect faint unresolved X-ray emission from the CSS radio
galaxy 3C~303.1. We detect a thermal component at kT $\simeq 0.8$ keV
which seems likely to be produced in the ISM of the host galaxy. There
is evidence for a second component in the spectrum whose nature is
currently ambiguous. It does not appear to be related to the core,
giving 3C 303.1 the advantage over other GPS/CSS radio galaxies that extended
components show up more clearly.
Plausible hypotheses
include (1) hot gas shocked by the expansion of the radio source, and
(2) SSC emission from the southern radio lobe if the magnetic field is
below the equipartition value by a factor of about 3.5.  
In the former case this would be the first detection of strongly
shocked gas around an FRII radio source.
In the latter case it would be a rare instance of the detection of SSC
emission from a radio lobe, as distinct from a small localized hot
spot.  Deep {\it Chandra\/} observations could confirm the presence of
this extended component and improve our understanding of its origin.


\acknowledgements

We would like to thank the referee for helpful comments and Joel Carvalho for 
helping to interpret the numerical simulations. 
Support for this work was provided by NASA through grant number NNG04GE70G.
This research made use of (1) the NASA/IPAC Extragalactic Database
(NED) which is operated by the Jet Propulsion Laboratory, California
Institute of Technology, under contract with the National Aeronautics and
Space Administration; and (2)  NASA's Astrophysics Data System Abstract
Service. 


\clearpage
\begin{deluxetable}{lr}
\tablewidth{0pt}
\tablecaption{Source Properties \label{tabsource}}
\tablehead{
\colhead{Parameter } &
\colhead{3C~303.1}
}
\startdata
ID &  G  \\ 
redshift &  0.267 \\
RA  (J2000) & 14:43:14.8 \\
DEC & 77:07:28 \\
scale (kpc/arcsec)  & 4.07 \\
radio power log$_{10}$P$_{ 5 \rm GHz}$ (Watts Hz\mone)  & 26.53 \\
radio source total angular size $\theta$ (arcsec) & 1.7 \\
linear size $D$ (kpc) & 6.9 \\
integrated emission line flux F(OIII$\lambda$5007) ($10^{-15}$ ergs s\mone\ cm\mtwo) &
 28 \\
spectral age (yr)  & $1\times 10^5$ \\
advance speed (v/c) & 0.07 \\
\enddata
\tablecomments{
We adopt the standard cosmology with  $H_o = 71$ km s\mone\ Mpc\mone,
$\Omega_{\rm m} = 0.3$, and $\Omega_\Lambda = 0.7$.
The integrated emission line flux is  from Gelderman \& Whittle (1994). 
The spectral age is estimated 
by fitting a continuous injection model to the integrated radio spectrum
and is taken from Murgia \etal\ (1999). The advance speed is estimated using 
2 v = linear size / spectral age. 
 }
\end{deluxetable}

\begin{deluxetable}{lccc}
\tablewidth{0pt}
\tablecaption{XMM Observations of 3C~303.1 \label{tabobs}}
\tablehead{
\colhead{Parameter } &
\colhead{pn} &
\colhead{Mos1} &
\colhead{Mos2} 
}
\startdata
Net Exposure Time (ksec) &  25.93 & 32.13 & 32.67  \\
Net Count Rate (counts/s) & 8.15E-3$\pm$ 1.92E-3 & 2.15E-3$\pm$ 7.22E-4 &
1.62E-3$\pm$ 6.97E-4 \\
Total Counts & 628 & 218 & 191 \\
Total Counts 0.4-8 keV & 483 & 127 & 128 \\
Net Counts$^a$ 0.2-12 keV  & 211 & 69 & 53 \\
\enddata
\tablecomments{
These numbers are after GTI filtering. Observation date is August, 18, 2003.
$^a$ background subtracted. 
 }
\end{deluxetable}

\begin{deluxetable}{lc}
\tablewidth{0pt}
\tablecaption{Spectral Fits: Single Thermal Component \label{tabtherm1}}
\tablehead{
\colhead{Parameter } &
\colhead{Value } 
}
\startdata
External Absorption $n_H$ ($\times 10^{22}$ cm\mtwo) & 0.031$^a$  \\
Abundance                                            & 1.0$^b$ \\
Temperature     kT (keV)                             & $0.82\pm 0.05$ \\
Luminosity 0.4-8.0 keV ($10^{42}$ ergs s\mone) & 2.4 \\
1 keV normalization  ($10^{-6}$ photons cm\mtwo\ sec\mone\ keV\mone) &
$7.7 \pm 2.1$ \\
Chi-squared & 24.63 \\
degrees of freedom & 27 \\
\enddata
\tablecomments{
A fit to wabs*raymond using the energy range 0.4 to 8.0 keV over 29 bins. 
$^a$Absorption is frozen at the foreground Galactic value.
$^b$Abundance is frozen at the solar value. 
 }
\end{deluxetable}

\begin{deluxetable}{lc}
\tablewidth{0pt}
\tablecaption{Spectral Fits: Thermal Component and Powerlaw \label{tabthermpow}}
\tablehead{
\colhead{Parameter } &
\colhead{Value }
}
\startdata
External Absorption $n_H$ ($\times 10^{22}$ cm\mtwo) & 0.031$^a$  \\
Abundance                                            & 1.0$^b$ \\
Temperature	kT (keV)                             & $0.77\pm 0.09$ \\
Luminosity 0.4-8.0 keV ($10^{42}$ ergs s\mone) & 2.0 \\
1 keV normalization  ($10^{-6}$ photons  cm\mtwo\ sec\mone\ keV\mone) & $5.9 \pm 4.3$ \\
Absorption towards the PL $n_H$ ($\times 10^{22}$ cm\mtwo) & $0.47 \pm 1.38$ \\
Photon index $\Gamma$ & 1.5$^c$ \\
1 keV PL normalization  ($10^{-6}$ photons cm\mtwo\ sec\mone\ keV\mone) &
$1.9^{0.1}_{-0.7}$ \\
Luminosity 0.4-8.0 keV ($10^{42}$ ergs s\mone) & 2.2 \\
Chi-squared & 17.17 \\
degrees of freedom & 25 \\
\enddata
\tablecomments{
A fit to wabs*(raymond + zwabs*pow) using the energy range 0.4 to 8.0 keV over 29 bins.
$^a$Absorption is frozen at the foreground Galactic value.
$^b$Abundance is frozen at the solar value.
$^c$The power-law photon index is poorly constrained by our data, so we froze a value
of $\Gamma = 1.5$. A value of $\Gamma = 2.0$ produces similar results. 
 }
\end{deluxetable}

\begin{deluxetable}{lc}
\tablewidth{0pt}
\tablecaption{Spectral Fits: Warm and Hot Thermal Components  \label{tabt2}}
\tablehead{
\colhead{Parameter } &
\colhead{Value }
}
\startdata
External Absorption $n_H$ ($\times 10^{22}$ cm\mtwo) & 0.031$^a$  \\
Abundance                                            & 1.0$^b$ \\
Temperature	kT (keV)                             & 0.8$^c$ \\
Luminosity 0.4-8.0 keV ($10^{42}$ ergs s\mone) & 1.7 \\
1 keV normalization  ($10^{-6}$ photons cm\mtwo\ sec\mone\ keV\mone) & $5.3 \pm 1.7$ \\
Temperature     kT (keV)                             & 45$^c$  \\
1 keV normalization  ($10^{-6}$ photons cm\mtwo\ sec\mone\ keV\mone) &
$13.6  \pm 7.6$ \\
Luminosity 0.4-8.0 keV ($10^{42}$ ergs s\mone) & 2.8 \\
Chi-squared & 18.77 \\
degrees of freedom & 27 \\
\enddata
\tablecomments{
A fit to wabs*(raymond + raymond) using the energy range 0.4 to 8.0 keV over 29 bins.
$^a$Absorption is frozen at the foreground Galactic value.
$^b$Abundance is frozen at the solar value.
$^c$Temperature frozen at 0.8 and 45 keV.
 }
\end{deluxetable}

\begin{deluxetable}{lcc}
\tablewidth{0pt}
\tablenum{6}
\tablecaption{Estimated Synchrotron Self-Compton Emission \label{tabssc}}
\tablehead{
\colhead{Parameter } &
\colhead{N Lobe } &
\colhead{S Lobe }
}
\startdata
Radio size (mas) & 440 x 200 & 180 x 80 \\
1.6 GHz radio flux density (mJy) & 140 & 1435 \\
radio spectral index $\alpha_r$ & 1.1 & 1.2 \\
minimum electron Lorentz factor $\gamma_{\rm min}$ & 1000 & 400 \\
Equipartition B field (milliGauss) & 0.17 & 0.7 \\
Predicted SSC X-ray 1 keV flux density (Jy)& $3.6\times 10^{-12}$ & $8.0\times 10^{-11}$ \\
B field correction factor & 0.06 & 0.29  \\
\enddata
\tablecomments{
The radio data are from 1.6 GHz MERLIN observations presented by Sanghera \etal\ (1995).
The B field correction factor is the factor needed to reduce the B field so that
the predicted SSC emission is consistent with the observed PL component. 
 }
\end{deluxetable}



\end{document}